\documentclass[twocolumn,showpacs,preprintnumbers,amsmath,amssymb]{revtex4}
\usepackage{graphicx}% Include figure files
\usepackage{dcolumn}% Align table columns on decimal point
\usepackage{bm}% bold math

\begin{document}

%\preprint{APS/123-QED}

\title{The discrepancy in the muon $g - 2$ is
a non-perturbative\\ effect of the Standard Model}% Force line breaks with \\
\author{Boris A. Arbuzov}
\affiliation{Skobeltsyn Institute of Nuclear Physics of Moscow State University\\
 Leninskie gory 1, 119991 Moscow Russia}
\email{arbuzov@theory.sinp.msu.ru}
\date{\today}% It is always \today, today,
\begin{abstract}
The non-perturbative approach based on Bogoliubov compensation principle is applied to the
calculation of a contribution to the muon $g-2$. Using the previous results on the spontaneous generation of the effective anomalous three-boson interaction the contribution is
calculated and shown to be in agreement with the well-known discrepancy. The calculated quantity contains no adjusting parameters but the experimental values for the muon and the W-boson masses.

\end{abstract}

%Use showkeys class option if keyword
\maketitle

Measurements of the anomalous magnetic moment of the muon $(g-2)_\mu$ (see the last publication~\cite{g-2ex} and the extensive review~\cite{Miller}) provides the only significant deviation of the experiment from predictions of the Standard Model. According to recent
analysis of the problem~\cite{g-2th1, g-2th2} this deviation $\Delta a_\mu$ safely exceeds four s.d. and comprises the following values correspondingly
\begin{eqnarray}
& &\Delta a_\mu\,=\,(3.493 \pm 0.823)\,10^{-9}\,;\label{amu}\\
& &\Delta a_\mu\,=\,(3.935 \pm 0.523_{th} \pm 0.63_{ex})\,10^{-9}\,.\nonumber
\end{eqnarray}

It should be emphasized, that the deviation from the SM calculations means
the deviation from perturbative calculations in the electro-weak theory. However there quite may be
non-perturbative contributions to physical quantities. In particular, the method of disclosing of the non-perturbative effects is developing starting of N.N. Bogoliubov compensation principle~\cite{NNB1, NNB2}.
In works~\cite{BAA04, BAA06, AVZ06, BAA07, AVZ09, BAA09,AZ11, AVZ2},
this principle
was applied to studies of a spontaneous generation of effective non-local interactions in renormalizable gauge theories of the Standard Model. For example, in works~\cite{BAA06, AVZ06} the well-known Nambu--Jona-Lasinio effective interaction was shown to be spontaneously generated in the framework of QCD. All the parameters of this interaction are calculated in terms of the initial QCD parameters.

In the present letter we apply the previous results to the problem of discrepancy $\Delta a_\mu$. It will come clear, that the effect under discussion is quite natural in the theory with account of the spontaneous generation of an effective interaction in the conventional electro-weak theory.

The main principle of the approach is to check if an effective interaction
could be generated in the chosen variant of a renormalizable theory. In view
of this one performs "add and subtract" procedure for the effective
interaction with a form-factor. Then one assumes the presence of the
effective interaction in the interaction Lagrangian and the same term with
the opposite sign is assigned to the newly defined free Lagrangian. The main point of the approach consists in formulation and solution of a compensation equation, which eliminates the undesirable interaction term from the new free Lagrangian. Provided this equation has a non-trivial solution the effective interaction is generated.

In works~\cite{AZ11, AVZ2}  the approach was applied to the electro-weak interaction and a possibility of spontaneous generation of anomalous three-boson interaction of the form
\begin{equation}
-\,\frac{G}{3!}\cdot\,\epsilon_{abc}\,W_{\mu\nu}^a\,W_{\nu\rho}^b\,W_{\rho\mu}^c\,;
\label{FFF}
\end{equation}
was studied. The notation~(\ref{FFF}) means corresponding
non-local vertex in the momentum space
\begin{eqnarray}
& &(2\pi)^4\,G\,\,\epsilon_{abc}\,(g_{\mu\nu} (q_\rho pk - p_\rho qk)+ g_{\nu\rho}
(k_\mu pq - q_\mu pk)+\nonumber\\
& &g_{\rho\mu} (p_\nu qk - k_\nu pq)+ q_\mu k_\nu p_\rho - k_\mu p_\nu q_\rho)\times\nonumber\\
& &F(p,q,k)\,\delta(p+q+k)+...;\label{vertex}
\end{eqnarray}
where $F(p,q,k)$ is a form-factor and
$p,\mu, a;\;q,\nu, b;\;k,\rho, c$ are respectfully incoming momenta,
Lorentz indices and weak isotopic indices
of $W$-bosons. We mean also that there are present four-boson, five-boson and
six-boson vertices according to the well-known non-linear expression for $W_{\mu\nu}^a$. Note, that in the approximation used we  maintain the gauge invariance of the approach.

Effective interaction~(\ref{FFF}) is
usually called anomalous three-boson interaction and it is considered for long time on phenomenological grounds~\cite{Hag}. Note, that the first attempt to obtain the anomalous three-boson interaction in the framework of Bogoliubov approach was done in work~\cite{Arb92}. The interaction constant $G$ is connected with
conventional definitions in the following way
\begin{equation}
G\,=\,-\,\frac{g\,\lambda}{M_W^2}\,.\label{Glam}
\end{equation}
The current limitations for parameter $\lambda$ read~\cite{EW, EW13}
\begin{eqnarray}
& &\lambda\,=\,-\,0.016^{+0.021}_{-0.023}\,;\; -\,0.059< \lambda < 0.026\,
(95\%\,C.L.)\,.
\nonumber\\
& &\lambda_\gamma\,=\,-\,0.022\,\pm 0.019\,;\label{EW13}
\end{eqnarray}
where the last number~(\ref{EW13}) is obtained recently by joint analysis of LEP data by the four experimental groups: ALEPH, DELPHI, L3, OPAL.
There is no difference in anomalous interaction for $Z$ and $\gamma$, i.e.
$\lambda_Z\,=\,\lambda_\gamma\,=\,\lambda$ according to standard relation
$W^0=\sin\theta_W A + \cos\theta_W Z$.

In works~\cite{AZ11, AVZ2}  the possibility of spontaneous generation of anomalous three-boson interaction of the form~(\ref{FFF})
was shown in the first approximation, corresponding to one-loop equation. In  the course of the study the following simple dependence of form-factor $F$ on all three variables was used
\begin{equation}
F(p_1,\,p_2,\,p_3)\,=\,F\Bigl(\frac{p_1^2\,+\,p_2^2\,+\,p_3^2}{2}\Bigr)\,;\label{123}
\end{equation}
Let us present the expression for four-boson vertex
\begin{eqnarray}
& &\frac{V(p,m,\lambda;\,q,n,\sigma;\,k,r,\tau;\,l,s,\pi)}{\imath\,(2 \pi)^4} =\nonumber\\
& &g\, G \biggl(\epsilon^{amn}
\epsilon^{ars}\Bigl(U(k,l;\sigma,\tau,\pi,\lambda)-\nonumber\\
& &U(k,l;\lambda,\tau,\pi,\sigma)-
U(l,k;\sigma,\pi,\tau,\lambda)+\nonumber\\
& &U(l,k;\lambda,\pi,\tau,\sigma)+U(p,q;\pi,\lambda,\sigma,\tau)-\nonumber\\
& &U(p,q;\tau,\lambda,\sigma,\pi)-U(q,p;\pi,\sigma,\lambda,\tau)+\nonumber\\
& &U(q,p;\tau,\sigma,\lambda,\pi)\Bigr)-\epsilon^{arn}\,
\epsilon^{ams}\times\nonumber\\
& &\Bigl(U(p,l;\sigma,\lambda,\pi,\tau)-U(l,p;\sigma,\pi,\lambda,\tau)-
\nonumber\\
& &U(p,l;\tau,\lambda,\pi,\sigma)+U(l,p;\tau,\pi,\lambda,\sigma)+
\label{four}\\
& &U(k,q;\pi,\tau,\sigma,\lambda)-U(q,k;\pi,\sigma,\tau,\lambda)-
\nonumber\\
& &U(k,q;\lambda,\tau,\sigma,\pi)+U(q,k;\lambda,\sigma,\tau,\pi)\Bigr)+
\nonumber\\
& &\epsilon^{asn}\,
\epsilon^{amr}\Bigl(U(k,p;\sigma,\lambda,\tau,\pi)-\nonumber\\
& &U(p,k;\sigma,\tau,\lambda,\pi)
+U(p,k;\pi,\tau,\lambda,\sigma)-\nonumber\\
& &U(k,p;\pi,\lambda,\tau,\sigma)-U(l,q;\lambda,\pi,\sigma,\tau)+\nonumber\\
& &U(l,q;\tau,\pi,\sigma,\lambda)-U(q,l;\tau,\sigma,\pi,\lambda)+\nonumber\\
& &U(q,l;\lambda,\sigma,\pi,\tau)\Bigr)\biggr);\nonumber\\
& &U(k,l;\sigma,\tau,\pi,\lambda)=(k_\sigma l_\tau g_{\pi\lambda}-k_\sigma\,
l_\lambda g_{\pi\tau}+\nonumber\\
& &k_\pi\,l_\lambda\,g_{\sigma\tau}-(kl)g_{\sigma\tau}g_{\pi\lambda}) F(k,l,-(k+l))\,.\nonumber
\end{eqnarray}

Here triad $p,\,m,\,\lambda$ {\it etc} means correspondingly momentum, isotopic index, Lorentz index of a boson; $g$ is the usual gauge coupling constant of the electro-weak interaction. Note, that in vertex~(\ref{four}) only momenta of two legs are present in various combinations. Thus a leg here is either "momentum" one or "sterile" one. We shall use these notations in what follows while discussing distribution of momenta in diagrams.

Now in the way of studying the problem~\cite{AZ11, AVZ2}
we get convinced, that there exists a non-trivial solution, which is expressed in terms of Meijer functions~\cite{BE}
$$
G_{qp}^{nm}\Bigl( z\,|^{a_1,..., a_q}_{b_1,..., b_p}\Bigr)\,;
$$
In case $q=0$ we write only indices $b_i$ in one line.
The solution for the form-factor is unique and has the the following form for $0 <z < z_0$
\begin{eqnarray}
& &F(z)\,=\,\frac{1}{2}\,G_{15}^{31}\Bigl( z\,|^0_{1,\,1/2,\,0,\,-1/2,\,-1}
\Bigr) -\nonumber\\
& &\frac{85\,g \sqrt{2}}{512\,\pi}\,G_{15}^{31}\Bigl( z\,|^{1/2}_{1,\,1/2,
\,1/2,\,-1/2,\,-1}\Bigr)\,+\nonumber\\
& &+\,C_1\,G_{04}^{10}\Bigl( z\,|1/2,\,1,\,-1/2,\,-1\Bigr)\,+
\label{solutiong}\\
& &C_2\,G_{04}^{10}\Bigl( z\,|1,\,1/2,\,-1/2,\,-1\Bigr)\,.\nonumber\\
& &z\,=\,\frac{G^2\,x^2}{512 \pi^2}\,;\quad x\,=\,p^2\,.
\nonumber
\end{eqnarray}
For $ z \geq z_0$ we have the trivial solution
\begin{equation}
F(z)\,=\,0\,.\label{F0}
\end{equation}
Parameters of solution~(\ref{solutiong}) are the following
\begin{eqnarray}
& &g\,=\,g(z_0)\,=\,0.60366\,;\quad z_0\,=\,9.61750\,;\nonumber\\
& &C_1\,=\,-\,0.035096\,;\quad
 C_2\,=\,-\,0.051104\,.\label{gY}
\end{eqnarray}

We would draw attention to the fixed value of parameter $z_0$. The solution
exists only for this value~(\ref{gY}) and it plays the role of eigenvalue.
As a matter of fact, from the beginning the existence of such eigenvalue is
by no means evident. The definite value for $g(z_0)$  is also worth mentioning. Let us explicate, that $g(z_0)$ is the value of running gauge coupling $g$ at the momentum $Q_0$, which is defined by relation $G^2\,Q^4_0 = 512\,\pi^2\,z_0$.

Emphasize, that an existence of a non-trivial solution of a compensation
equation is extremely
restrictive. In the most cases such solutions do not exist at all. When we start from a
renormalizable theory we have arbitrary value for
its coupling constant. Provided there exists
non-trivial solution of a compensation equation the
coupling is fixed as well as the parameters of this
non-trivial solution.

\begin{figure}
\includegraphics[width=5cm]{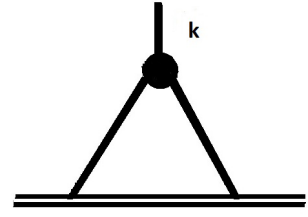}
\caption {One loop diagram for calculation of new contribution to the muon magnetic moment. Vertical line represents the photon, simple lines -- W bosons, black spot -- triple vertex~(\ref{FFF}) with corresponding form-factor. Double line represents the muon.}
\label{Fig1}
\end{figure}

\begin{widetext}

\begin{figure}
\includegraphics[width=18cm]{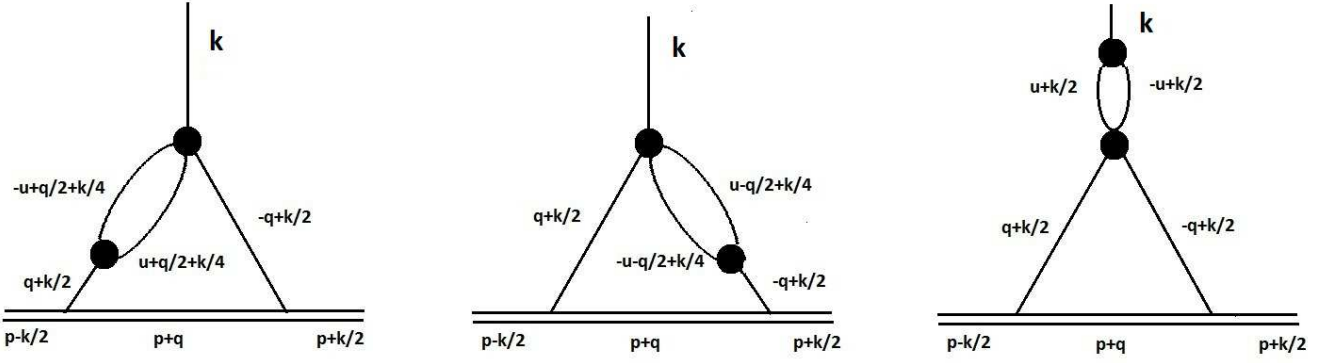}

\caption{Two loop diagrams for calculation of new contribution to the muon magnetic moment. Vertical line represents the photon, simple lines -- W bosons, black spots -- triple vertex~(\ref{FFF}) and four leg vertex~(\ref{four}) with corresponding form-factors. Double line represents the muon.The momenta are directed downwards and rightwards (the muon). The momentum of integration in the oval-like loops is denoted $u$.}
\label{Fig2}
\end{figure}
\end{widetext}

Now let us consider a contribution of interaction~(\ref{FFF}) with form-factor defined by relations~(\ref{123},\ref{solutiong},\ref{gY}) to the anomalous magnetic moment of the charged spin one half particle with mass $m$, for example, of the muon. The first approximation described by the simplest diagram presented in Fig.\ref{Fig1} gives zero. This result is immediately connected with Lorentz structure of anomalous vertex~(\ref{vertex}).

Thus to obtain non-zero contribution to $a_\mu$ we look for two-loop diagrams, which give contribution to three-boson Lorentz structure of the usual gauge vertex.
\begin{equation}
V_{p,\mu; q,\nu; k,\rho} = g_{\mu\nu}(q - p)_\rho + g_{\nu\rho}(k-q)_\mu +
g_{\rho\mu}(p-k)_\nu .\label{gauge}
\end{equation}
Diagrams shown in Fig.\ref{Fig2}, in which four-boson vertex corresponds to terms in expression~(\ref{four}) with "momentum" legs entering to oval loops achieve this goal.

The calculations are performed in the unitary gauge. The main contribution to the result is given by increasing terms in nominators in vector boson propagators, containing $M_W^{-2}$
$$
\frac{g_{\mu \nu}-\frac{q_\mu q_\nu}{M_W^2}}{q^2 -M^2_W}\,;
$$
while in denominators we put $M_W = 0$. The estimate of accuracy of this procedure will be given below.

In the course of calculations finite renormalization of the gauge coupling $g$ is performed. In doing this we single out the constant contribution to the Lorentz structure~(\ref{gauge}) of triple boson vertices including oval-like loops. After this procedure only the two first diagrams in Fig.(\ref{Fig2}) give contribution to the value of the magnetic moment. Remind, that the structure of the anomalous vertex~(\ref{FFF}) gives zero contribution.

We have for the main contribution to a magnetic moment according to diagrams Fig. \ref{Fig2} the following Lorentz structure
\begin{eqnarray}
& &\frac{\hat p\; \hat k\, \gamma_a}{4}\,-\,\frac{\gamma_a\, \hat k\, \hat p}{4}\,-\,\frac{1}{3}\hat p\,\gamma_a\, \hat p\,=\nonumber\\
& &\frac{m}{6}\,\frac{\hat k\,\gamma_a\,-\,\gamma_a\,\hat k}{2}\,;\label{sigma}
\end{eqnarray}
which is multiplied by the two-loop integral, which we calculate in the Euclid four dimensional momentum space. The integration momentum inside an oval loop is $u$ and inside a triangle loop is $q$. Denoting $u^2 = x$ and $q^2 = y$ we have from~(\ref{123}) combination $x + \frac{3 y}{4}$ for arguments of both form-factors. Thus substitution $t = x + \frac{3 y}{4}$ is quite natural. Using variables $t$ and $y$ we have the following expression for coefficient before the magnetic moment structure with account of~(\ref{sigma})
\begin{eqnarray}
& &\frac{ m\, e\, g^2 G^2}{12 (16 \pi^2)^2 M_W^2}\int_0^Y dt F^2(t)\biggl(\int_0^t \frac{4\, t \, dy}{(6 t-3 y)}+\nonumber\\
& &\int_t^{4 t/3}\frac{4\, t (16 t^3-48 t^2 y+48 t y^2-15 y^3)\,dy}{3(2 t-y) y^3}\biggr);\label{da1}
\end{eqnarray}
where $Y$ is defined by the relation
$$
z_0\,=\,\frac{G^2\,Y^2}{512\,\pi^2}\,.
$$
From~(\ref{da1}) with definitions of variable $z$ and of the form-factor~(\ref{solutiong}, \ref{gY}) we obtain the following result for the contribution to $a_\mu$
\begin{eqnarray}
& &\Delta a_\mu\,=\,\frac{g(z_0)^2\,m^2}{3\,\pi^2\,M_W^2}\,\biggl(20\,\ln\biggl[\frac{4}{3}\biggr]
-\frac{13}{3}\biggr)\times\nonumber\\
& &\int_0^{z_0}\, F^2(z)\,dz\,=\,2.775\times10^{-9}\,.\label{int2}
\end{eqnarray}
where we have used only values of the muon mass and the $W$-boson mass.
All other parameters are defined by solution~(\ref{solutiong}) with parameters~(\ref{gY}). Let us draw attention to the disappearance of the effective interaction coupling constant $G$ from expression~(\ref{int2}). This is due to entering of factor $G^2$
into the denominator according to definition of variable $z$. Thus the main result does not depend on $\lambda$. This parameter influence only the next approximations.
Let us estimate possible corrections due to
$M_W \neq 0$. They are defined by the following parameter
\begin{equation}
\frac{\sqrt{2} g |\lambda|}{32 \pi}\,=\,0.0005\,;
\end{equation}
with the maximal value of $|\lambda|=0.059$ from restrictions~(\ref{EW13}). Thus this correction is negligible. The other correction may be connected with the value of gauge coupling $g(z_0)$. It is taken from solution~(\ref{gY}). However it is possible to calculate experimental value for this parameter. Let us start from the well-known expression for the running electro-weak coupling with the total number of flavors
\begin{eqnarray}
& &g^2(Q^2)=\frac{g^2}{1+\frac{5\,g^2}{24 \pi^2}\,\ln\Bigl[\frac{Q^2}{M_W^2}\Bigr]}\,;\label{gQ}\\
& &g = g(M_W^2) = 0.65\,;\;Q^2(z_0) = \frac{32\, \pi \sqrt{z_0}\,M_W^2}{\sqrt{2}\, g\, |\lambda|}\nonumber
\end{eqnarray}
Then with the same $|\lambda|=0.059$ we obtain $g(z_0) = 0.626$ and with this
value we have instead of result~(\ref{int2})
\begin{equation}
\Delta a_\mu = 2.987\times 10^{-9}\,.\label{int3}
\end{equation}
This value is few {\it per cent} larger than value~(\ref{int2}). There may be also other corrections to result~(\ref{int2}). The examples being studied earlier have given estimate for accuracy of the approximation $\simeq 10\%$~\cite{BAA04, AVZ06}. Bearing in mind this estimate, the result for the non-perturbative contribution to $a_\mu$ is advertised to be the following
\begin{equation}
\Delta a_\mu\,=\,(2.78\,\pm 0.28)\times10^{-9}\,.\label{result}
\end{equation}
The result, as well as values~(\ref{int2}, \ref{int3}), evidently agrees deviation~(\ref{amu}) within error bars.

There are proposals to connect the $\Delta a_\mu$ effect with theories beyond the Standard Model, for example with effects of SUSY~\cite{SUSY}. However with such proposals one inevitably introduces additional parameters to adjust the discrepancy. Here we have no adjusting parameters. Therefore the result~(\ref{result}) is to be considered as an evidence for confirmation of the Standard Model. What is necessary, is to learn how to calculate non-perturbative contributions. The method, which is used in the present note, gives quite an adequate result.
Thus we consider the result
to be encouraging and promising to further applications of
the Bogoliubov compensation approach to principal problems
of elementary particles physics.

The author express gratitude to I.V. Zaitsev for valuable discussions.

The work was supported in part by grant of Ministry of education and science of RF No. 8412 and by grant NSh-3920.2012.2.

\end{document}